\documentclass[12pt]{article}
\def\c2x2{c$(2\times 2)$}
\def\2x1{$(2\times 1)$}

\def\avg#1{\left\langle{#1}\right\rangle}
\def\dn2n{{\avg{\left(\delta N\right)^2}\over \avg{N}}}
\def\tdn2n{{\avg{\left(\delta N\right)^2} / \avg{N}}}

\begin{document}
\renewcommand{\baselinestretch}{1.3}\small\normalsize
\begin{center}

{\Large\bf A CLASS OF STOCHASTIC AND DISTRIBUTIONS-FREE QUANTUM MECHANICS EVOLUTION EQUATIONS: \\ \vspace{.3cm} THE SCHR\"{O}DINGER-LIKE EQUATIONS}

\vspace{2cm}
\renewcommand{\baselinestretch}{1.0}\small\normalsize

{G. Costanza}

\vspace{1cm}

{Departamento de F\'{i}sica, Instituto de F\'{i}sica Aplicada (INFAP),
Universidad Nacional de San Luis, Chacabuco 917, 5700
San Luis, Argentina}

\vspace{0.6cm}

\end{center}
\vspace{1cm}

\begin{abstract}
A procedure allowing to construct rigorously discrete as well as continuum deterministic evolution equations from stochastic evolution equations is developed using a Dirac's bra and ket notation. This procedure is an extension of an approach previously used  by the authors coined Discrete Stochastic Evolution Equations. Definitions and examples of discretes as well as continuum one-dimensional lattices are developed in detail in order to show the basic tools that allows to construct Schr\"{o}dinger-like equations. Extension to $d$-dimensional lattices are studied in order to provide a wider exposition and the one-dimensional cases are derived as special cases, as expected. Some variants of the procedure allows to construct other evolution equations. Also, using a limiting procedure, it is possible to derive the Schr\"{o}dinger equations from the Schr\"{o}dinger-like equations.
\end{abstract}

\vspace{5cm}

\renewcommand{\baselinestretch}{1.3}\small\normalsize
\newpage

\section{\label{sec:level1}Introduction}
In this paper, a deterministic quantum mechanics evolution equation is derived from a set of stochastic quantum mechanics evolution equation using a bra and ket notation which can be considered an extension of an approach coined Discrete Stochastic Evolution Equations (DSEE) [1]. Some illustrative examples that show the versatility of this approach can be found in [1-9]. Within this context, discrete as well as continuum Schr\"{o}dinger-like equations will be obtained without the use of distributions like Dirac's delta. This goal is achieved using an appropriate split of the discrete as well as the continuum Hamiltonian, allowing a complete rigorous derivation of the usual equation proposed or obtained in the literature, e.g. [9,10,11]. Lot of work was done and is doing on the subject of Quantum Stochastic Processes, see for example the classical books [12,13] or more recently from a more mathematical grounds e.g. [14-16]. Appropriate expansions in a finite centered differences series as well as in a Taylor series for the discrete and continuum evolution equation, respectively, allows to complete the basic tools used for deriving distribution-free equations in a straightforward way. The possibility of obtaining deterministic evolution equations from stochastic evolution equations is based on the assumption that the Hamiltonian and the wave functions are statistically independent. This means that the averages over realization of products of the Hamiltonian with the wave functions factorizes. One more assumption that allows the obtention of discrete as well as continuum Schr\"{o}dinger-like equations is that the Hamiltonian can be split in a sum of two terms, as will be seen in all the examples given below. An explanation about the use of Schr\"{o}dinger-like equations instead of Schr\"{o}dinger equation is in order. It is well known (see e.g. [11] chapter 16) that the Hamiltonian allowing to obtain a deterministic one-dimensional Schr\"{o}dinger equation is expressed in terms of distributions like

\begin{eqnarray}
H(x,x') &=& -\frac{\hbar}{2m}\delta''(x-x')+V(x)\delta(x-x')\nonumber\\&=& \left(-\frac{\hbar}{2m}\frac{d^2}{dx^2}+V(x)\right)\delta(x-x') ,\nonumber
\end{eqnarray}

where $x$ and $x'$ are two points, $\delta''(x-x')$ is the second derivative of the delta function $\delta(x-x')$, $V(x)$ is the potential,  $m$ is the mass of the particle, and $\hbar$ is the reduced Plank constant. On the other hand, it will be referred here to Schr\"{o}dinger-like equation to an equation with the same form that the Schr\"{o}dinger equation when the Hamiltonian and the wave function are expanded in a Taylor series up to an appropriate order in $\triangle x'=x'-x$, as we will seen below. The question it will be answered is: Is it possible to find a Hamiltonian, independent of distributions, that after introducing it in the general evolution equations

\begin{eqnarray}
\frac{d \psi_i(t)}{d t}&=&-\frac{\imath}{\hbar}\sum_jH_{ij}(t)\psi_j(t),\hspace{2cm}  \forall i,j \in \Lambda_1,\nonumber
\end{eqnarray}

or

\begin{eqnarray}
\frac{\partial \psi(x,t)}{\partial t}&=&-\frac{\imath}{\hbar}\int H(x,x',t)\psi(x',t)dx' ,\hspace{.5cm}  \forall x,x' \in \Re,\nonumber
\end{eqnarray}

 generates a discrete or continuum Schr\"{o}dinger-like equation, respectively? Note that the Hamiltonian can also be time dependent as it will be see below. The answer is yes for both discrete as well as continuum cases. The Schr\"{o}dinger equation can be obtained using a limiting procedure.

The paper is organized as follows. In Section 2, the simplest case is considered and the basic definitions of a stochastic one-dimensional discrete evolution equation of a ket is used in order to find a discrete deterministic evolution equation for the wave function and the corresponding discrete Schr\"{o}dinger-like equations. This approach allows to prove that the usual proposed evolution equations can be obtained rigorously from first principle. In Section 3, the basic definitions of a stochastic one-dimensional evolution equation of a ket is used in order to find a continuum deterministic evolution equation for the wave function and the corresponding continuum Schr\"{o}dinger-like equation. In Section 4, the extension to a $d$-dimensional lattices are considered and a continuum Schr\"{o}dinger-like equation is obtained and the one-dimensional case is obtained as a special case, as expected. In Section 5, some additional illustrative examples are considered in order to provide a better understanding of the procedures. In Section 6 it is proved that the Schr\"{o}dinger equations can be obtained as a limiting case of a Schr\"{o}dinger-like equations. In Section 7, conclusions, some generalizations, and perspectives are given .

\section{The discrete one-dimensional lattice}

Beginning, for the sake of simplicity in the presentation, with the simplest stochastic evolution of a one-dimensional ket, namely

\begin{eqnarray}
 \mid\psi^{(r)}(t+\Delta t)\rangle &=& U^{(r)}(t+\Delta t,t)\mid\psi^{(r)}(t)\rangle \nonumber\\&=& \left(1- \frac{\imath}{\hbar}H'^{(r)}(t)\Delta t+O(\Delta t^2) \right)\mid\psi^{(r)}(t)\rangle \nonumber\\&=& \mid \psi^{(r)}(t)\rangle - \frac{\imath}{\hbar}H'^{(r)}(t)\Delta t\mid\psi^{(r)}(t)\rangle,
\end{eqnarray}

where $(r)$ indicates that the evolution refers to a particular realization and a "discrete" Taylor series expansion of  $U^{(r)}(t+\Delta t,t)$ up to $O(\Delta t)$  and $U^{(r)}(t,t)=1$, was used. In this way, an evolution equation similar to the one for the dynamical variables given in [9] is obtained for the corresponding ket $\mid\psi^{(r)}(t)\rangle$. The so called "weights" in [9], in this case, are $\frac{\Delta  U^{(r)}(t,t)}{\Delta t}\Delta t=- \frac{\imath}{\hbar}H'^{(r)}(t)\Delta t$, where $\imath=\surd\overline{-1}$, $\hbar$ is the reduced Plank constant and $H'^{(r)}(t)$ is the "primed Hamiltonian" which will be the Hamiltonian when $ \Delta  t \rightarrow 0 $. As usual, if the evolution equations for the probability amplitudes is required, it must be done the following operations on both two sides of Eq.(1): 1) multiplying by a bra $\langle i|$, 2) summing $-\langle i\mid \psi^{(r)}(t)\rangle$ , 3) dividing  by  $ \Delta  t $ , and 4) letting $ \Delta  t \rightarrow 0 $. After these steps are completed, $\frac{d  U^{(r)}(t,t)}{dt}=- \frac{\imath}{\hbar}H^{(r)}(t)$ and the following evolution equation is obtained

\begin{eqnarray}
\frac{d \psi_i^{(r)}(t)}{d t}&=&\langle i|-\frac{\imath}{\hbar}H^{(r)}(t)\mid\psi^{(r)}(t)\rangle\nonumber\\&=&-\frac{\imath}{\hbar}\langle i|H^{(r)}(t) \sum_j \mid j \rangle \langle j\mid\psi^{(r)}(t)\rangle \nonumber\\&=&-\frac{\imath}{\hbar}\sum_j\langle i|H^{(r)}(t)  \mid j \rangle \langle j\mid\psi^{(r)}(t)\rangle\nonumber\\&=&-\frac{\imath}{\hbar}\sum_jH_{ij}^{(r)}(t)\psi_j^{(r)}(t),\hspace{.5cm}  \forall i,j \in \Lambda_1,
\end{eqnarray}

where $\Lambda_1$ is the set of sites of the lattice with periodic boundary conditions, and

\begin{eqnarray}
\frac{d \psi_i^{(r)}(t)}{d t}&=&\lim_{\Delta t \rightarrow 0 }\frac{\psi_i^{(r)}(t+\Delta t)- \psi_i^{(r)}(t)}{\Delta t} ,\nonumber\\\mid\psi^{(r)}(t)\rangle&=&\sum_j \mid j \rangle \langle j\mid\psi^{(r)}(t)\rangle, \nonumber\\ H^{(r)}_{ij}(t)&=& \langle i|H^{(r)}(t)  \mid j \rangle,  \nonumber\\  \psi_i^{(r)}(t)&=&\langle i\mid\psi^{(r)}(t)\rangle,\nonumber\\ \psi_j^{(r)}(t)&=&\langle j\mid\psi^{(r)}(t)\rangle,\nonumber
\end{eqnarray}

was used. Note that it was used $i$, $j$ as integer number but this discrete sites are particular cases of $a_1 i$ and $a_1 j$ corresponding to the coordinates of the lattice sites which they are equals only if  the spacing of the lattice $a_1$ is one. It must be emphasized that sometimes in the literature [9,11] the notation used is $\langle i\mid\psi^{(r)}(t)\rangle = C_i^{(r)}(t)$ instead of $\psi_i^{(r)}(t)$, but here it will be used the notation usually used in the literature. If a deterministic evolution equation is required, an average over realization on both two sides of Eq.(2) is needed. The final result is

\begin{eqnarray}
\frac{d \psi_i(t)}{d t}&=&-\frac{\imath}{\hbar}\sum_jH_{ij}(t)\psi_j(t),\hspace{1.2cm}  \forall i,j \in \Lambda_1,
\end{eqnarray}

where,

\begin{eqnarray}
 \psi_i(t)&=&\overline{\psi_i^{(r)}(t)},\nonumber\\ H_{ij}(t)&=&\overline{H_{ij}^{(r)}(t)},\nonumber\\ H_{ij}(t)\psi_j(t)&=&\overline{H_{ij}^{(r)}(t)\psi_j^{(r)}(t)}=\overline{H_{ij}^{(r)}}(t)\hspace{0.3cm}\overline{\psi_j^{(r)}(t)}.
\end{eqnarray}

It is easy to see that a factorization was assumed, which means that  $H_{ij}^{(r)}(t) $ and $\psi_j^{(r)}(t)$ are statistically independent. In other words, a "discrete" deterministic evolution equation as the one given in Eq.(3) can only be obtained if the stochastic Hamiltonian and the probability amplitudes are statistically independent. One way different to the ones considered in [9,11], where a discrete Scchr\"{o}dinger-like equation is obtained, is by making $j=i+\triangle j$ with $\triangle j=j-i$, and letting to split the Hamiltonian as $H_{ij}(t)=(H_1)_{ij}+(H_0)_{ii}\frac{\psi_i(t)}{\psi_j(t)}$, where $(H_1)_{ij}$ and $(H_0)_{ii}$ are both two time independent. Using these definitions, Eq.(3) becomes

\begin{eqnarray}
\frac{d \psi_i(t)}{d t}&=&-\frac{\imath}{\hbar}\left(\sum_j(H_1)_{ij}\psi_j(t)+\sum_j(H_0)_{ii}\psi_i(t)\right)\nonumber\\&=&-\frac{\imath}{\hbar}\left(S_1+S_0\right),
\end{eqnarray}

where

\begin{eqnarray}
S_1&=&\sum_j(H_1)_{ij}\psi_j(t)=\sum_j(H_1)_{ii+\triangle j}\psi_{i+\triangle j}(t) \nonumber\\&&\approx \sum_j(H_1)_{ii }\left(\psi_{i}(t)+\frac{\triangle\psi_{i}(t)}{\triangle i}\left(\triangle j\right)+ \frac{1}{2}\frac{\triangle^2\psi_{i}(t)}{\triangle i^2}\left(\triangle j\right)^2+\cdots\right),\nonumber\\& =&(H_1)_{ii }\sum_j\left(\psi_{i}(t)+\frac{\triangle\psi_{i}(t)}{\triangle i}\left(\triangle j\right)+ \frac{1}{2}\frac{\triangle^2\psi_{i}(t)}{\triangle i^2}\left(\triangle j\right)^2+\cdots\right) \nonumber\\& =& (H_1)_{ii }\left(\psi_{i}(t) s_0+\frac{\triangle\psi_{i}(t)}{\triangle i}s_1+ \frac{1}{2}\frac{\triangle^2\psi_{i}(t)}{\triangle i^2}s_2\right),\nonumber\\S_0&=&\sum_j(H_0)_{ii}\psi_i(t)=(H_0)_{ii}\psi_i(t)\sum_j(1)=(H_0)_{ii}\psi_i(t)s_0.
\end{eqnarray}

In $S_1$ it was used

\begin{eqnarray}
(H_1)_{ii+\triangle j}&=&(H_1)_{ii}+ 0(\triangle j),\nonumber\\ \psi_{i+ \triangle j}(t)&=&\psi_i(t)+\frac{\triangle \psi_i(t)}{\triangle i}\left(\triangle j\right) +\frac{1}{2}\frac{\triangle^2 \psi_i(t)}{\triangle i^2}\left(\triangle j\right)^2  \nonumber\\ &+& 0((\triangle j)^3) ,\nonumber
\end{eqnarray}

where, the finite central (or centered) differences used are $\triangle i= a_1$, $\triangle i^2= a_1^2$,  $\triangle\psi_{i}(t)=\psi_{i+\frac{1}{2}}(t)-\psi_{i-\frac{1}{2}}(t)$, and $\triangle^2\psi_{i}(t)=\triangle\left(\triangle\psi_{i}(t)\right)=\psi_{i-1}(t)-2\psi_{i}(t)+\psi_{i+1}(t)$.

The summations $s_0$, $s_1$, and $s_2$ are

\begin{eqnarray}
s_0&=&\sum_j(1)= \overbrace{1+\cdots+1}^{N_1 times}=  N_1\approx 2 N'_1,\nonumber\\s_1&=& \sum_j\left(\triangle j\right)=\left(-N'_1+\cdots+0+\cdots+N'_1\right)=0,\nonumber\\s_2&=& \sum_j\left(\triangle j\right)^2 =\left(-N'_1\right)^2+\cdots+\left(0\right)^2+\cdots+\left( N'_1\right)^2\nonumber\\& =&\frac{2}{3}\left(N'_1\right)^3 + \left(N'_1\right)^2+\frac{1}{3}\left(N'_1\right)\approx\frac{2}{3}\left(N'_1\right)^3 ,
\end{eqnarray}

where, if the number of lattice sites $N_1$ is odd and  $N_1\gg 1$, then $N'_1=\frac{N_1-1}{2}\approx \frac{N_1}{2}$.  Note that in Eq.(5), using this approximation, the time derivative of $\psi_i(t)$ depends only on the nearest neighbor  values like a continuum Schr\"{o}dinger-like equation where only second order partial derivative with respect to the spacial coordinate appears in the evolution equation, as will be shown below. Using the results of Eqs.(6,7) in Eq.(5), it is easy to find

\begin{eqnarray}
\frac{d \psi_i(t)}{d t}&=&-\frac{\imath}{\hbar}\left( s_2\frac{(H_1)_{ii}}{2a_1^2}\left[\psi_{i-1}(t)-2\psi_{i}(t)+\psi_{i+1}(t)\right] \right)\nonumber\\&-& \frac{\imath}{\hbar}s_0\left[ (H_0)_{ii}+(H_1)_{ii} \right]\psi_i(t)\nonumber\\&=&-\frac{\imath}{\hbar}s_2\frac{(H_1)_{ii}}{2a_1^2}\left[ \psi_{i-1}(t)+\psi_{i+1}(t)\right]\nonumber\\&-&\frac{\imath}{\hbar}(\triangle H)_{ii}\psi_i(t) ,\hspace{2cm}  \forall i \in \Lambda_1,
\end{eqnarray}

where $(\triangle H)_{ii}=s_0[(H_0)_{ii}+(H_1)_{ii}]-2\frac{(H_1)_{ii}}{2a_1^2}s_2=(H_0)_{ii}s_0+(H_1)_{ii}\left( s_0-\frac{s_2}{a_1^2}\right)$. Note that this discrete evolution equation is the same that the one proposed in [11], if $(H_1)_{ii}=-\frac{2a_1^2A}{s_2}$ and $(H_0)_{ii}=\frac{2a_1^2As_0+s_2E_0-2As_2}{s_0s_2}$. These values are obtained after equating each of the coefficients with the ones given in [11] and solving the set of linear equations

\begin{eqnarray}
(H_0)_{ii}s_0+(H_1)_{ii}\left( s_0-\frac{s_2}{a_1^2}\right)= E_0,\nonumber\\ s_2 \frac{(H_1)_{ii}}{2a_1^2}=-A,\nonumber
\end{eqnarray}

for $(H_1)_{ii}$ and $(H_0)_{ii}$.

 The meaning of the coefficients in [11] are: $E_0$ is a constant that allows to choose the zero of the energy, $A$ is a constant independent of $t$, and $b=a_1$ is the spacing of the lattice. Of course, here Eq.(8) was derived from first principles.\\

\section{The continuum one-dimensional lattice}

The way of obtaining a continuum Schr\"{o}dinger-like equation is to transform Eq.(2,3) making the transformations

\begin{eqnarray}
\psi_i^{(r)}(t)&\rightarrow& \psi^{(r)}(x,t)=\langle x\mid\psi^{(r)}(t)\rangle ,\nonumber\\\psi_j^{(r)}(t)&\rightarrow& \psi^{(r)}(x',t)=\langle x'\mid\psi^{(r)}(t)\rangle  ,\nonumber\\ H^{(r)}_{ij}(t) &\rightarrow& H^{(r)}(x,x',t)=\langle x\mid H^{(r)}(t)\mid x'\rangle \nonumber,
\end{eqnarray}

where the discrete indices $i$ and $j$ where replaced by the continuous variables $x$ and $x'$, respectively. After transforming the summation into an integral like

\begin{eqnarray}
-\frac{\imath}{\hbar}\sum_jH_{ij}^{(r)}(t)\psi_j^{(r)}(t)&\rightarrow& -\frac{\imath}{\hbar}\int H^{(r)}(x,x',t)\psi^{(r)}(x',t)dx',\nonumber
\end{eqnarray}

where $dx'$ is the differential length, $H^{(r)}(x,x',t)$ is the Hamiltonian density or the Hamiltonian per unit of length (or, in general, $d$-dimensional volume), and $\psi^{(r)}(x,t)$ is the wave function. \textbf{It must be warning that even when the same symbol was used for the discrete and continuum Hamiltonian, they are not the same. Also hereafter hamiltonian will be used for both the discrete and continuum case}. Moreover, in a $d$-dimensional lattice, the differential volume is  $dV'_d=dx'_1...dx'_d$, and the integral, as usual, is a multiple integral, one per dimension as it will be shown below. The final result is

\begin{eqnarray}
\frac{\partial \psi^{(r)}(x,t)}{\partial t}&=&-\frac{\imath}{\hbar}\int H^{(r)}(x,x',t)\psi^{(r)}(x',t)dx' ,\nonumber\\&& \hspace{3.4cm}  \forall x,x' \in \Re,
\end{eqnarray}

and after an average over realization

\begin{eqnarray}
\frac{\partial \psi(x,t)}{\partial t}&=&-\frac{\imath}{\hbar}\int H(x,x',t)\psi(x',t) dx' \nonumber\\&& \hspace{3.4cm}  \forall x,x' \in \Re,
\end{eqnarray}

where it was assumed that $H^{(r)}(x,x',t)$ and $\psi^{(r)}(x',t)$ are statistically independent, consequently, $\overline{H^{(r)}(x,x',t)\psi^{(r)}(x',t)}=\overline{H^{(r)}(x,x',t)}\hspace{.3cm} \overline{\psi^{(r)}(x',t)}= H(x,x',t)\psi(x',t)$. If it is needed to obtain a Schr\"{o}dinger-like equation, the Hamiltonian must be split in a sum like  $H(x,x',t)=H_1(x,x')+H_0(x,x)\frac{\psi(x,t)}{\psi(x',t)}$, where $H_0(x,x)$ and $H_1(x,x)$  are both two time independent. Then Eq.(10) becomes

\begin{eqnarray}
\frac{\partial \psi(x,t)}{\partial t}&=& -\frac{\imath}{\hbar}\int H_1(x,x')\psi(x',t) dx'\nonumber\\&-&\frac{\imath}{\hbar}I_1 H_0(x,x)\psi(x,t),
\end{eqnarray}

where, making  $dx'=d\left(\triangle x'\right)$ with  $\triangle x'=x'-x$(see Eqs.(14,15) below),

\begin{eqnarray}
I_1&=& \int_{-L'_1}^{L'_1} d(\triangle x')=2L'_1.\nonumber
\end{eqnarray}

In order to obtain a Taylor series expansion of the right hand side of Eq.(11), it is necessary to rewrite the evolution equation in a convenient way in order to obtain

\begin{eqnarray}
 \psi(x',t)&=& \psi(x+\triangle x',t)= \psi(x,t)+ \frac{\partial \psi(x,t)}{\partial x}(\triangle x')\nonumber\\&+& \frac{1}{2} \frac{\partial^2 \psi(x,t)}{\partial x^2}(\triangle x')^2+O((\triangle x')^3) \nonumber\\H_1(x,x')&=& H_1(x,x+\triangle x')= H_1(x,x)+ O((\triangle x')).\nonumber\\
\end{eqnarray}

After replacing Eq.(12) in Eq.(11) and integrating over $x'$ it is easily found
\begin{eqnarray}
\frac{\partial \psi(x,t)}{\partial t}&=&-\frac{\imath}{\hbar}\left(h_1+h_2+ h_3+h_0\right)+ O((\triangle x')^3),\nonumber\\
\end{eqnarray}

where
\begin{eqnarray}
h_1&=&H_1(x,x)\psi(x,t) I_1\nonumber\\h_2&=& H_1(x,x)\frac{\partial \psi(x,t)}{\partial x} I_2\nonumber\\h_3&=& H_1(x,x)\frac{1}{2} \frac{\partial^2 \psi(x,t)}{\partial x^2}I_3\nonumber\\h_0&=& I_1 H_0(x,x)\psi(x,t),\nonumber
\end{eqnarray}

and

\begin{eqnarray}
I_1&=& \int_{x-L'_1}^{x+L'_1} dx'=\int_{-L'_1}^{L'_1} d(\triangle x')=2L'_1,\nonumber\\I_2&=& \int_{-L'_1}^{L'_1} (\triangle x') d(\triangle x')=\left[\frac{(\triangle x')^2}{2}\right ]_{-L'_1}^{L'_1}=0,\nonumber\\I_3&=& \int_{-L'_1}^{L'_1} (\triangle x')^2 d(\triangle x')=\left[\frac{(\triangle x')^3}{3}\right ]_{-L'_1}^{L'_1}=2\left[\frac{(L'_1)^3}{3}\right ].\nonumber\\
\end{eqnarray}

Note that these integrals are the same as the summations $s_0$, $s_1$, and $s_2$, in Eq.(7), except for the fact that half the number of points of the lattice $N_1$ is here halve the length of the lattice $L_1$. In Eq.(14) it was used the following change of variable

\begin{eqnarray}
d(\triangle x')&=&\frac{d(x'-x)}{d x'}dx'=dx'.
\end{eqnarray}

Letting

\begin{eqnarray}
\left[ H_0(x,x)+H_1(x,x) \right] I_1&=&V(x),\nonumber\\\frac{1}{2}H_1(x,x)I_3&=&\left(\frac{-\hbar^2}{2m_{eff}}\right),
\end{eqnarray}

in Eq.(13), the usual Schr\"{o}dinger-like equation is obtained. Note that it is possible to choose $H_0(x,x)$ , $H_1(x,x)$, and $L_1$, in such a way that the results obtained from a discrete lattice, given in [9,11], are recovered. The final result is

\begin{eqnarray}
H_0(x,x)&=& \frac{E_0}{I_1},\nonumber\\ H_1(x,x)&=& -\frac{2 A}{I_3},\nonumber\\L'_1&=& b \surd\overline{3}.
\end{eqnarray}

 It is not difficult to see that, even when the formal results obtained in Eq.(16) are correct, in the last equation, in Eq.(17), the value of halve the length of the lattice, obtained after equating $m_{eff}=\left(\frac{\hbar^2}{2A b^2}\right)$, given in [11], to  $m_{eff}=\left(\frac{-\hbar^2}{H_1(x,x)I_3}\right)$, is too small and consequently a meaningless result. Let to provide a "reasonable physical assumption" that allows to obtain an alternative appropriate result. In order to achieve this goal, let to assume that the Hamiltonian is different from zero only inside an interval $-L''_1\leq \triangle x'\leq L''_1$ with $|L''_1|< L_1$, then the integrals in Eq.(14) are

\begin{eqnarray}
I'_1&=& \int_{x-L''_1}^{x+L''_1} dx'=\int_{-L''_1}^{L''_1} d(\triangle x')=2L''_1,\nonumber\\I'_2&=& \int_{-L''_1}^{L''_1} (\triangle x') d(\triangle x')=\left[\frac{(\triangle x')^2}{2}\right ]_{-L''_1}^{L''_1}=0,\nonumber\\I'_3&=& \int_{-L''_1}^{L''_1} (\triangle x')^2 d(\triangle x')=\left[\frac{(\triangle x')^3}{3}\right ]_{-L''_1}^{L''_1}=2\left[\frac{(L''_1)^3}{3}\right ],\nonumber\\
\end{eqnarray}

and consequently, Eq.(16) becomes

\begin{eqnarray}
\left[ H'_0(x,x)+H'_1(x,x) \right] I'_1&=&V(x),\nonumber\\\frac{1}{2}H'_1(x,x)I'_3&=&\left(\frac{-\hbar^2}{2m_{eff}}\right),
\end{eqnarray}

and Eq.(17) must be changed to

\begin{eqnarray}
H'_0(x,x)&=& \frac{E_0}{I'_1},\nonumber\\ H'_1(x,x)&=& -\frac{2 A}{I'_3},\nonumber\\L''_1&=& b\surd\overline{3},
\end{eqnarray}

where $L''_1$ was obtained after equating $m_{eff}=\left(\frac{\hbar^2}{2A b^2}\right)$ to  $m_{eff}=\left(\frac{-\hbar^2}{H'_1(x,x)I'_3}\right)$. This sort of "cut off" provide an interval which is very much small than the length of the lattice $L_1$ and is very close to the lattice spacing $b=a_1$, which is a very reasonable physical result and justify the approximate value of the Hamiltonian in Eq.(12).

\section{Extensions to a $d$-dimensional lattice}

The extension to a $d$-dimensional lattice is straightforward. Defining $d$-dimensional vectors $\overrightarrow{x}$ and $\overrightarrow{x}'$, the Taylor series expansion of the wave function is

\begin{eqnarray}
 \psi(\overrightarrow{x'},t)&=& \psi(\overrightarrow{x}+\triangle \overrightarrow{x}',t)= \psi(\overrightarrow{x},t)+ \sum_{k=1}^{\infty}\frac{1}{k!} \nabla_d^k \psi(\overrightarrow{x},t)\nonumber\\
\end{eqnarray}

where $k!$ is the factorial of $k$, $\triangle\overrightarrow{x}'=\overrightarrow{x}'-\overrightarrow{x}$, and

\begin{eqnarray}
 \nabla_d= \left( \triangle x'_1 \frac{\partial}{\partial x_1}+ \cdots  +\triangle x'_d \frac{\partial}{\partial x_d}\right),
\end{eqnarray}

where $\triangle x'_l=x'_l-x_l$, for $l=1,\cdots,d$, is the $l$-th component of $\triangle\overrightarrow{x}'$. As usual, in Eq.(21)

\begin{eqnarray}
 \psi(\overrightarrow{x}',t)&=& \psi(x'_1,\cdots,x'_d,t),\nonumber\\ \psi(\overrightarrow{x}+\triangle \overrightarrow{x}',t)&=& \psi(x_1+\triangle x'_1,\cdots,x_d+\triangle x'_d,t).\nonumber\\
\end{eqnarray}

With the above extension to a $d$-dimensional lattice it is possible to rewrite all the one-dimensional equations. For example,

\begin{eqnarray}
\frac{d \psi_{i_1 \cdots i_d}(t)}{d t}&=&-\frac{\imath}{\hbar}\sum_{j_1} \cdots \sum_{j_d}H_{i_1\cdots i_d,j_1 \cdots j_d}(t)\psi_{j_1 \cdots j_d}(t), \nonumber\\ &&\hspace{1cm}\forall i_1,\cdots ,i_d,j_1,\cdots ,j_d \in \Lambda_d,
\end{eqnarray}

where $i_1, \cdots ,i_d,j_1, \cdots ,j_d $ are the components of the sites $i,j$ in a $\Lambda_d$ lattice. Using a finite $d$-dimensional centered differences, extension of the discrete one-dimensional case can be obtained. The explicit form will be not considered here and only the continuum version will be analyzed. To this end, the definitions corresponding to the $d$-dimensional extensions are

\begin{eqnarray}
\frac{\partial \psi(\overrightarrow{x},t)}{\partial t}&=&-\frac{\imath}{\hbar}\int H(\overrightarrow{x},\overrightarrow{x}',t)\psi(\overrightarrow{x}',t) dV'\nonumber\\&=& -\frac{\imath}{\hbar}\int H_1(\overrightarrow{x},\overrightarrow{x}')\psi(\overrightarrow{x}',t) dV'\nonumber\\&-&\frac{\imath}{\hbar}I_0 H_0(\overrightarrow{x},\overrightarrow{x})\psi(\overrightarrow{x},t),\nonumber\\&& \hspace{.6cm}  \forall x_1,\cdots,x_d,x'_1,\cdots,x'_d \in \Re,
\end{eqnarray}

where, as in the one-dimensional case, $H(\overrightarrow{x},\overrightarrow{x}',t)=H_1(\overrightarrow{x},\overrightarrow{x}')+H_0(\overrightarrow{x},\overrightarrow{x}) \frac{\psi(\overrightarrow{x},t)}{\psi(\overrightarrow{x}',t)}$, with $H_1(\overrightarrow{x},\overrightarrow{x}')$ and $H_0(\overrightarrow{x},\overrightarrow{x}')$ are both two time independent, and

\begin{eqnarray}
I_0&=& \int_{-L'_1}^{L'_1}\cdots \int_{-L'_d}^{L'_d}d\left(\triangle x'_1\right) \cdots d\left(\triangle x'_d\right))=\left(2L'_1\right)\cdots\left(2L'_d\right). \nonumber
\end{eqnarray}

 Using the extension of the change of variables given in Eq.(15), it is easy to obtain

\begin{eqnarray}
d(\triangle V')&=&d(x_1'-x_1) \cdots d(x_d'-x_d)=d\left(\triangle x'_l\right) \cdots d\left(\triangle x'_d\right) \nonumber\\&=&\frac{d(x_1'-x_1)}{d x_1'}d x_1' \cdots\frac{d(x_d'-x_d)}{d x_d'} dx_d'= dx_1'\cdots dx_d'\nonumber\\&=& dV'.
\end{eqnarray}

The final $d$-dimensional evolution equation can be written in the following convenient form

\begin{eqnarray}
\frac{\partial \psi(\overrightarrow{x},t)}{\partial t}&=&-\frac{\imath}{\hbar}\int_{-L'_1}^{L'_1}\cdots \int_{-L'_d}^{L'_d} H_1(\overrightarrow{x},\overrightarrow{x}')\psi(\overrightarrow{x}',t)\mathcal{D}\triangle '\nonumber\\ &+& H_{00} \hspace{.5cm}  \forall x_1,\cdots,x_d,x'_1,\cdots,x'_d \in \Re,
\end{eqnarray}

where $H_{00}=-\frac{\imath}{\hbar}I_0 H_0(\overrightarrow{x},\overrightarrow{x})\psi(\overrightarrow{x},t)$. It was also used $d(\triangle V')=d\left(\triangle x'_l\right) \cdots d\left(\triangle x'_d\right)=\mathcal{D}\triangle '$, in the second equality, which is very convenient when the expansion of $\psi(\overrightarrow{x}+\triangle\overrightarrow{x}',t)$ and $H_1(\overrightarrow{x},\overrightarrow{x}+\triangle\overrightarrow{x}')$, in a Taylor series, provides a set of elementary integrations as it was done in Eqs.(18) for a one-dimensional lattice. Finally, using the Taylor series expansion given in Eq.(21) up to $k=2$, and letting $H_1(\overrightarrow{x},\overrightarrow{x}+\triangle\overrightarrow{x}')=H_1(\overrightarrow{x},\overrightarrow{x})+0(\triangle \overrightarrow{ x}')$, a $d$-dimensional Schr\"{o}dinger-like equation is obtained. As in the one-dimensional case, if only three term are kept, it is easy to find

\begin{eqnarray}
\frac{\partial \psi(\overrightarrow{x},t)}{\partial t}&=& H_{00}+\int_{-L'_1}^{L'_1}\cdots \int_{-L'_d}^{L'_d}H_{11}\left(1+ \nabla_d +\frac{\nabla_d^2}{2}\right)\nonumber\\ &\times & \psi(\overrightarrow{x},t)\mathcal{D}\triangle ' \nonumber\\ &=& H_{00}+H_{11}I_0\psi(\overrightarrow{x},t)+ I_1+I_2,\nonumber\\ && \hspace{1cm}  \forall x_1,\cdots,x_d,x'_1,\cdots,x'_d \in \Re,
\end{eqnarray}

where $H_{11}=-\frac{\imath}{\hbar}H_1(\overrightarrow{x},\overrightarrow{x})$,

\begin{eqnarray}
I_1&=& -\frac{\imath}{\hbar}H_1(\overrightarrow{x},\overrightarrow{x})\int_{-L'_1}^{L'_1}\cdots \int_{-L'_d}^{L'_d}\nabla \psi(\overrightarrow{x},t)\mathcal{D}\triangle ' \nonumber\\&=&H_{11}\int_{-L'_1}^{L'_1}\cdots \int_{-L'_d}^{L'_d}\left(\triangle x'_1 \frac{\partial}{\partial x_1}+ \cdots  +\triangle x'_d \frac{\partial}{\partial x_d}\right)\nonumber\\&\times&\psi(\overrightarrow{x},t)\mathcal{D}\triangle ' \nonumber\\&=&H_{11}\left(I_{1,1} \frac{\partial\psi(\overrightarrow{x},t)}{\partial x_1}+ \cdots  +I_{1,d} \frac{\partial\psi(\overrightarrow{x},t)}{\partial x_d}\right),\nonumber\\
\end{eqnarray}

and

\begin{eqnarray}
I_2&=& -\frac{\imath}{\hbar}H_1(\overrightarrow{x},\overrightarrow{x})\int_{-L'_1}^{L'_1}\cdots \int_{-L'_d}^{L'_d}\frac{\nabla_d^2}{2} \psi(\overrightarrow{x},t)\mathcal{D}\triangle '  \nonumber\\&=&\frac{H_{11}}{2}\int_{-L'_1}^{L'_1}\cdots \int_{-L'_d}^{L'_d}\left( \triangle x'_1 \frac{\partial}{\partial x_1}+ \cdots  +\triangle x'_d \frac{\partial}{\partial x_d}\right)^2 \nonumber\\&\times& \psi(\overrightarrow{x},t)\mathcal{D}\triangle '.\nonumber\\
\end{eqnarray}

 It is not difficult to see that all the integrals on the righthand side of Eq.(29) are zero. Remembering that $\frac{\partial\psi(\overrightarrow{x},t)}{\partial x_l}=constant$ $\forall l =1,\cdots,d$, $I_1$ is a sum of elementary integrals. A generic multiple integral corresponding to the $l$-th term is a product of integrals like

\begin{eqnarray}
I_{1,l}&=& \int_{-L'_1}^{L'_1}\cdots \int_{-L'_d}^{L'_d}\triangle x'_l d\left(\triangle x'_1\right) \cdots d\left(\triangle x'_d\right))\nonumber\\&=& I'_1\cdots I'_l\cdots I'_d=0,\hspace{.3cm}  \forall l=1,\cdots,d,
\end{eqnarray}

where

\begin{eqnarray}
 I'_{1}&=& \triangle x'_l \int_{-L'_1}^{L'_1} d\triangle x'_1=\triangle x'_l \left[\triangle x'_1)\right ]_{-L'_1}^{L'_1}= 2L'_1\triangle x'_l ,\nonumber\\ \vdots\nonumber\\I'_{l}&=& \int_{-L'_l}^{L'_l}\triangle x'_l d\triangle x'_l=\left[\frac{(\triangle x'_l)^2}{2}\right ]_{-L'_1}^{L'_1}=0,\nonumber\\ &&\hspace{.1cm} \forall l=1,\cdots,d, \nonumber\\ \vdots\nonumber\\I'_{d}&=& \triangle x'_l \int_{-L'_d}^{L'_d} d\triangle x'_d=\triangle x'_l \left[\triangle x'_d)\right ]_{-L'_d}^{L'_d}= 2L'_d\triangle x'_l .\nonumber\\
\end{eqnarray}

Due that always one of the factors, in each of the $d$ terms in Eq.(29), is zero because $I'_{l}=0$ $\forall l=1,\cdots,d$, then $I_{1,l}=0$ $\forall l=1,\cdots,d$ and consequently also $I_1=0$. After expanding the righthand side in Eq.(30)

\begin{eqnarray}
I_{2}&=&\frac{H_{11}}{2}\sum_{l.l'}\left(\int_{-L'_1}^{L'_1}\cdots \int_{-L'_d}^{L'_d} \triangle x'_l \triangle x'_{l'}\frac{\partial^2\psi(\overrightarrow{x},t)}{\partial x_l\partial x_{l'}}\mathcal{D}\triangle'\right),\nonumber\\&=&\frac{H_{11}}{2}\sum_{l.l'}\frac{\partial^2\psi(\overrightarrow{x},t)}{\partial x_l\partial x_{l'}}\left(\int_{-L'_1}^{L'_1}\cdots \int_{-L'_d}^{L'_d} \triangle x'_l \triangle x'_{l'}\mathcal{D}\triangle'\right),\nonumber\\&=&\frac{H_{11}}{2}\sum_{l.l'}\frac{\partial^2\psi(\overrightarrow{x},t)}{\partial x_l\partial x_{l'}}\left(I_{l,l'}\right)
,
\end{eqnarray}

where

\begin{eqnarray}
I_{l,l'}&=& \int_{-L'_1}^{L'_1}\cdots \int_{-L'_d}^{L'_d}\triangle x'_l\triangle x'_{l'} d\left(\triangle x'_1\right)\cdots d\left(\triangle x'_d\right)\nonumber\\&=& I'_1\cdots I'_l\cdots I'_d=0,\hspace{1cm}\forall l'\neq l,
\end{eqnarray}

because

\begin{eqnarray}
I'_{l'}&=& \int_{-L'_l}^{L'_l}\triangle x'_l\triangle x'_{l'} d\left(\triangle x'_l\right)= \triangle x'_{l'}\int_{-L'_l}^{L'_l}\triangle x'_l d\left(\triangle x'_l\right)\nonumber\\&=& \triangle x'_{l'} \left[\frac{\triangle x'^2_{l}}{2} \right ]^{L'_l}_{-L'_1}=0, \hspace{1cm}  \forall l'\neq l.
\end{eqnarray}

On the other hand

\begin{eqnarray}
I'_{l'}&=& \int_{-L'_l}^{L'_l}\triangle x'_l\triangle x'_{l'} d\left(\triangle x'_l\right)= \int_{-L'_l}^{L'_l}\triangle x'^2_l d\left(\triangle x'_l\right)\nonumber\\&=&  \left[\frac{\triangle x'^3_{l}}{3} \right ]^{L'_l}_{-L'_1}= 2\frac{L'^3_{l}}{3} ,\hspace{1cm}  \forall l'= l,
\end{eqnarray}

consequently

\begin{eqnarray}
I_{l,l}&=& I'_{l'}\overbrace{\int_{-L'_1}^{L'_1}\cdots \int_{-L'_d}^{L'_d} }^{\left(d-1\right) times}d\left(\triangle x'_1\right)\cdots d\left(\triangle x'_d\right)\nonumber\\&=& 2\frac{L'^3_{l}}{3}\left( \prod _{k\neq l}2L'_k \right) ,
\end{eqnarray}

where $I_{l,l'}= I_{l,l}$ was used, indicating that only those terms where $l'=l$ remains. The other $d-1$ integrals are all of them of the form

\begin{eqnarray}
I_{k}&=& \int_{-L'_k}^{L'_k} d\left(\triangle x'_k\right)=  2L'_k , \hspace{1cm}  \forall k\neq l.\nonumber
\end{eqnarray}

The final result is

\begin{eqnarray}
I_{2}&=& \frac{H_{11}}{2}\sum_{l.l}\frac{\partial^2\psi(\overrightarrow{x},t)}{\partial x_l\partial x_l}\left(I_{l,l}\right)
,\nonumber\\&=& \frac{H_{11}}{2}\sum_{l}\frac{\partial^2\psi(\overrightarrow{x},t)}{\partial x_l^2}\left(I_{l,l}\right).
\end{eqnarray}

Assuming $L'_1=\cdots=L'_d=L'$ then $I_{l,l}=2\frac{L'^3}{3}\left( 2L'\right)^{d-1}$ $\forall l=1, \cdots ,d$  and consequently

\begin{eqnarray}
I_{2}&=& \frac{H_{11}}{2}\left(I_{l,l}\right)\sum_{l}\frac{\partial^2\psi(\overrightarrow{x},t)}{\partial x_l^2},\nonumber\\&=&\frac{H_{11}}{2}\left(2\frac{L'^3}{3}\right)\left( 2L'\right)^{d-1}\nabla^2\psi(\overrightarrow{x},t),\nonumber\\&=&-\frac{\imath}{\hbar}H_1(\overrightarrow{x},\overrightarrow{x})\left(\frac{L'^3}{3}\right)\left( 2L'\right)^{d-1}\nabla^2\psi(\overrightarrow{x},t),
\end{eqnarray}

where

\begin{eqnarray}
\nabla^2&=& \frac{\partial^2\psi(\overrightarrow{x},t)}{\partial x_1^2}+\cdots+\frac{\partial^2\psi(\overrightarrow{x},t)}{\partial x_d^2}.\nonumber
\end{eqnarray}

It is not difficult to see that for $d=1$ the one-dimensional results are recovered. In order to connect the above evolution equation with a continuum Schr\"{o}dinger-like equation it is necessary to make

\begin{eqnarray}
\left[ H_0(\overrightarrow{x},\overrightarrow{x})+H_1(\overrightarrow{x},\overrightarrow{x}) \right]I_0 &=&V(\overrightarrow{x}),\nonumber\\ H_1(\overrightarrow{x},\overrightarrow{x})\left(\frac{L'^3}{3}\right)\left(2L'\right)^{d-1}&=& \left(\frac{-\hbar^2}{2m_{eff}}\right),
\end{eqnarray}

where $I_0=\left(2L'_1\right)\cdots\left(2L'_d\right)=\left(2L'\right)^d$. With this results Eq.(27) becomes

\begin{eqnarray}
\frac{\partial \psi(\overrightarrow{x},t)}{\partial t}&=&-\frac{\imath}{\hbar}\left( -\frac{\hbar^2}{2M_{eff}} \nabla^2 \psi(\overrightarrow{x},t)+V(\overrightarrow{x})\psi(\overrightarrow{x},t)\right),\nonumber
\end{eqnarray}

which is the well known deterministic Schr\"{o}dinger-like equation.

As expected, for $d=1$ the results of Eqs.(16,17) or  Eqs.(19,20), depending of the limits of the integrals, are recovered.

\section{Other uses and extensions}

Now it will be considered other possibilities of the present approach.

\subsection{evolution equation like the heat equation}

For example, choosing the right hand side of Eq.(40) in other different form, it could be find another evolution equation like the heat equation. Letting

\begin{eqnarray}
\left[ H_0(\overrightarrow{x},\overrightarrow{x})+H_1(\overrightarrow{x},\overrightarrow{x}) \right]I_0 &=&\mu \left(\frac{-\hbar}{\imath}\right) ,\nonumber\\ H_1(\overrightarrow{x},\overrightarrow{x})\left(\frac{L'^3}{3}\right)\left(2L'\right)^{d-1}&=& \alpha \left(\frac{-\hbar}{\imath}\right),
\end{eqnarray}

where $\alpha$ and $\mu$ are real coefficients, solving for $H_0(\overrightarrow{x},\overrightarrow{x})$ and $H_1(\overrightarrow{x},\overrightarrow{x})$ and after replace both two in Eq.(27), it is possible to find the following evolution equation

\begin{eqnarray}
\frac{\partial \psi(\overrightarrow{x},t)}{\partial t}&=&\alpha \nabla^2 \psi(\overrightarrow{x},t)+\mu\psi(\overrightarrow{x},t).
\end{eqnarray}

\subsection{Inclusion of nonlinear terms }

Inclusion of nonlinear terms are possible letting

\begin{eqnarray}
H(\overrightarrow{x},\overrightarrow{x}',t) &=&\left[ H_0(\overrightarrow{x},\overrightarrow{x}',t) \frac{\psi(\overrightarrow{x},t)}{\psi(\overrightarrow{x'},t)}+H_1(\overrightarrow{x},\overrightarrow{x}',t) \right]\nonumber\\&=&\left[ H_0(\overrightarrow{x},\overrightarrow{x},t) \frac{\psi(\overrightarrow{x},t)}{\psi(\overrightarrow{x}',t)}+H_1(\overrightarrow{x},\overrightarrow{x},t) \right].\nonumber\\
\end{eqnarray}

Both $ H_0(\overrightarrow{x},\overrightarrow{x},t)$ and $ H_1(\overrightarrow{x},\overrightarrow{x},t)$ are now time dependent Hamiltonians and were obtained after taking the Taylor series expansion up to the first term like $H_0(\overrightarrow{x},\overrightarrow{x}',t)=H_0(\overrightarrow{x},\overrightarrow{x},t)+0(\triangle \overrightarrow{ x}')$  and $H_1(\overrightarrow{x},\overrightarrow{x}',t)=H_1(\overrightarrow{x},\overrightarrow{x},t)+0(\triangle \overrightarrow{ x}')$.With this modifications both two sides in Eq.(40) becomes

\begin{eqnarray}
\left[ H_0(\overrightarrow{x},\overrightarrow{x},t)+H_1(\overrightarrow{x},\overrightarrow{x},t) \right]I_0 &=&\kappa_0\psi(\overrightarrow{x},t) \psi\ast(\overrightarrow{x},t)\nonumber\\ &=&\kappa_0|\psi(\overrightarrow{x},t)|^2,\nonumber\\ H_1(\overrightarrow{x},\overrightarrow{x},t)\left(\frac{L'^3}{3}\right)\left(2L'\right)^{d-1}&=&\kappa_2,
\end{eqnarray}

where $\kappa_0$ and $\kappa_2$ are constants. Then, solving for $H_0(\overrightarrow{x},\overrightarrow{x},t)$ and $H_1(\overrightarrow{x},\overrightarrow{x},t)$ and after replace both two in Eq.(27), it is possible to find the following nonlinear deterministic Schr\"{o}dinger equation

\begin{eqnarray}
\frac{\partial \psi(\overrightarrow{x},t)}{\partial t}&=&-\frac{\imath}{\hbar}\left(\kappa_2\nabla^2 \psi(\overrightarrow{x},t)+\kappa_0|\psi(\overrightarrow{x},t)|^2\psi(\overrightarrow{x},t)\right).\nonumber\\
\end{eqnarray}

Of course, lot of other possibilities can be obtained by changing the right hand side of Eq.(40) or Eq.(45).

\subsection{Extension to a system of many particles}

The extension to a many particles system is straightforward. The coordinates, the wave function, and the Laplacian, must be changed to

\begin{eqnarray}
\overrightarrow{x}&=&\overrightarrow{x}_1,\cdots,\overrightarrow{x}_N,\nonumber\\
\psi(\overrightarrow{x},t)&=&\psi(\overrightarrow{x}_1,\cdots,\overrightarrow{x}_N,t),\nonumber\\
V(\overrightarrow{x},t)&=&V(\overrightarrow{x}_1,\cdots,\overrightarrow{x}_N,t),\nonumber\\
\nabla^2 &=&\sum^N_{k=1} \nabla^2_k ,\nonumber
\end{eqnarray}

where $N$ is the number of particles, $\overrightarrow{x}_k=x_{1,k},\cdots,x_{d,k}$ are the coordinates of the $k$-th.  particle, and $ \nabla^2_k = \frac{\partial^2}{\partial x^2_{1,k}}+ \cdots+ \frac{\partial^2}{\partial x^2_{d,k}}$, for $k=1,\cdots,N$.

\section{The Schr\"{o}dinger equation as a limiting case of the Schr\"{o}dinger-like equation}

It is possible to find the Schr\"{o}dinger equation as a particular case of the Schr\"{o}dinger-like equation as follows. Beginning with Eq.(13) and taking into account all the remaining terms in the Taylor series expansion of $ \psi(x',t)$ in Eq.(12), it is possible to prove that all the remaining terms are zero, for a particular value of the limits of the integral. The remaining terms are

\begin{eqnarray}
 R(x',t)&=& H'_1(x,x)\int_{-L''_1}^{L''_1} \sum _{n=3}^\infty \frac{\partial^n \psi(x,t)}{\partial x^n}\frac{(\triangle x')^n}{n!} d (\triangle x').\nonumber\\&=& \sum _{n=3}^\infty  \frac{\partial^n \psi(x,t)}{\partial x^n} I_n(L''_1),
\end{eqnarray}

where

\begin{eqnarray}
 I_n(L''_1)&=& H'_1(x,x)\int_{-L''_1}^{L''_1}\frac{(\triangle x')^n}{n!} d (\triangle x')\nonumber\\&=&\frac{2H'_1(x,x)}{n!}  \frac{(L''_1)^{n+1}}{n+1}.
\end{eqnarray}

and it wa used  $L''_1$ instead of $L'_1$ in order to avoid misunderstandings, like it was made in Eq.(18), because here again it will be assumed that the Hamiltonian is different from zero in an interval which will be choosing in order to make the remaining term zero.

Using $H'_1(x,x)$, given in Eq.(16), the final result is

\begin{eqnarray}
 I_n(L''_1)&=&\frac{2}{n!} \frac{C}{(L''_1)^{3}} \frac{(L''_1)^{n+1}}{n+1}=C_n (L''_1)^{n-2},
\end{eqnarray}

where

\begin{eqnarray}
C&=&\frac{-3\hbar^2}{2m_{eff}},\nonumber\\ C_n&=&\frac{2C}{n!(n+1)}.
\end{eqnarray}

It is obvious that for $L''_1\rightarrow 0$, $ R(x',t)\rightarrow 0$ in Eq.(46), then the Schr\"{o}dinger-like equation becomes a Schr\"{o}dinger equation. The extension to the $d-$dimensional case, using again $L''_1$ instead of $L'_1$, is straightforward. Beginning with the remainder terms of Eq.(28)

\begin{eqnarray}
R(\overrightarrow{x'},t)&=& \int_{-L''_1}^{L''_1}\cdots \int_{-L''_d}^{L''_d}H_{11}\left(\sum_{ n=3}^{\infty } \frac{\nabla_d^n}{n!}\right) \psi(\overrightarrow{x},t)\mathcal{D}\triangle ' \nonumber\\ &=& H_{11}\left(\sum_{ n=3}^{\infty }\partial^n I_n \right)=\sum_{ n=3}^{\infty }\partial^n H_{11}I_n,
\end{eqnarray}

where

\begin{eqnarray}
\partial^n &=& \frac{1}{n!} \sum_{ k_1=1}^{n}\cdots\sum_{ k_n=1}^{n} \frac{\partial^n\psi(\overrightarrow{x},t)}{\partial x_{k_1}\cdots \partial x_{k_n}},
\end{eqnarray}

\begin{eqnarray}
I_n&=& \int_{-L''_1}^{L''_1}\cdots \int_{-L''_d}^{L''_d} \triangle'_{x_{k_1}}\cdots \triangle'_{x_{k_n}}\mathcal{D}\triangle ' ,
\end{eqnarray}

and

\begin{eqnarray}
\mathcal{D}\triangle '&=& d \left( \triangle'_{x_{k_1}}\right)\cdots d \left( \triangle'_{x_{k_n}}\right) .
\end{eqnarray}

Using similar arguments that the ones that lead to Eq.(37) the multiple integral, assuming that $L''_1=\cdots=L''_d$, is

\begin{eqnarray}
I_n&=& 2\frac{L''^{(n+1)}_{1}}{n+1}\left( 2L''_1 \right)^{(d-1)} ,
\end{eqnarray}

and using the value of $H_1(\overrightarrow{x},\overrightarrow{x})$, obtained from Eq.(40), it is easy to see that

\begin{eqnarray}
R_n&=&H_{11}I_n\nonumber\\ &=& C_d \left( \frac{L''^{3}_{1}}{3}\left( 2L''_1 \right)^{(d-1)} \right)^{-1} 2\frac{L''^{(n+1)}_{1}}{n+1} \left( 2L''_1 \right)^{(d-1)} \nonumber\\&=& 6C_d\frac{L''^{(n-2)}_{1}}{n+1},
\end{eqnarray}

where $C_d$ is a constant independent of $L''_1$ . It is not difficult to see that for $L''_1\rightarrow 0$, $R_n\rightarrow 0\hspace{.5cm} \forall n\geq 3$ and consequently $R(\overrightarrow{x'},t)\rightarrow 0$. This result prove that the Schr\"{o}dinger equations can be obtained from a Schr\"{o}dinger-like equations as a limiting case.

\section{Conclusions}

The present paper deals with the obtention of deterministic Schr\"{o}dinger-like equations from stochastic evolution equations after an average over realization both for discrete as well as continuum equations. The deterministic version can be compared to some previous results, where the proposed equations in [11], here are rigorously obtained from first principles. Also the Schr\"{o}dinger equations can be obtained as a particular case of a Schr\"{o}dinger-like equations using a limiting procedure, as it was done in Section 6. Many other evolution equations can be obtained by simply change the right hand side of Eq.(44) and in the most general form like

\begin{eqnarray}
\left[ H_0(\overrightarrow{x},\overrightarrow{x},t)+H_1(\overrightarrow{x},\overrightarrow{x},t) \right]I_0 &=&F_0(\overrightarrow{x},t),\nonumber\\ H_1(\overrightarrow{x},\overrightarrow{x},t)\left(\frac{L'^3}{3}\right)\left(2L'\right)^{d-1}&=& F_2(\overrightarrow{x},t),
\end{eqnarray}
generates the nonlinear evolution equation
\begin{eqnarray}
\frac{\partial \psi(\overrightarrow{x},t)}{\partial t}&=&-\frac{\imath}{\hbar}\left(F_2(\overrightarrow{x},t)\nabla^2 \psi(\overrightarrow{x},t)+F_0(\overrightarrow{x},t)\psi(\overrightarrow{x},t)\right),\nonumber
\end{eqnarray}

  where $F_2(\overrightarrow{x},t)$ and $F_0(\overrightarrow{x},t)$ are arbitrary complex functions allowing to find all previous results, and many others, as special cases (see e.g.[16] for other examples of nonlinearity). It is easy to see that the new equations, in Eq.(56), are in general a system of four linear equations allowing to find the real and imaginary part of $H_1(\overrightarrow{x},t)$ and $H_0(\overrightarrow{x},t)$. Of course, this is a pure formal extension and the meaning, in each case, must be discussed. Note that all the discrete and continuum Schr\"{o}dinger-like equations where obtained essentially using three main steps. Firstly, the assumption that the stochastic hamiltonians and the wave functions are statistically independent, secondly, the split of the Hamiltonian in two terms like $H(\overrightarrow{x},\overrightarrow{x}',t)=H_0(\overrightarrow{x},\overrightarrow{x}',t)\frac{ \psi(\overrightarrow{x},t)}{\psi(\overrightarrow{x}',t)}+H_1(\overrightarrow{x},\overrightarrow{x}',t)$ and, thirdly, the expansion in a series up to an appropriate given order of $\triangle x'_l=x_l'-x_l $ for $l=1,\cdots, d$.

  Lot of additional extensions can be worked as, for example, the evolution equation of operators. This task will be considered in some future work.
\section{Acknowledgments}

The authors want to thank V. D. Pereyra for reading the manuscript.\\

\newpage

\section{References}

\begin{itemize}

\item[[1]] E. F. Costanza and G. Costanza, Physica A, {\bf 459} (2016)32.

\item[[2]] G. Costanza, Physica A, {\bf 388} (2009)2600.

 \item[[3]]G. Costanza, Physica A, {\bf 390} (2011)1713.

\item[[4]] G. Costanza, Physica A, {\bf 390} (2011)2267.

\item[[5]]  G. Costanza, Physica A, {\bf 391} (2012)2167.

\item[[6]]G. Costanza, Rev. Mex. Fis., {\bf 59} (2013)141.

\item[[7]] G. Costanza, Physica A {\bf 402} (2014)224.

 \item[[8]]  G. Costanza, Surf. Sci.{\bf 601} (2007)2095.

\item[[9]] E. F. Costanza and G. Costanza, Physica A, {\bf 435} (2015)51 and references therein.

\item[[10]] R. P. Feynman,  A. R. Hibbs, Quantum Mechanics and Path Integral, McGraw-Hill, New York,1965.

\item[[11]] R. P. Feynman, R. B. Leighton, and M. Sands, The Feynman Lectures on Physics, Quantum Mechanics Vol. III, Addison-Wesley, 1965 (Chapters 13,16)

\item[[12]]  N. G.Van Kampen, "Stochastic processes in Physics and Chemestry", Elsevier, 1997.

\item[[13]] C. W. Gardiner, "Handbook of Stochastic Methods", Springer, 1997.

\item[[14]] E. L. Hudson and K. R. Parthasarathy, Commun. Math. Physi.  {\bf 93} (1984)301.

\item[[15]] L. Accardi, F. Fagnola, and J. Quaegebeur, Journal of Functional Analisis,{\bf 104}(1992)149. Ricardo Castro Santis and Alberto Barchielly, arXiv:1001.2826v1.

\item[[16]] Miloslav Znojil, Franti$\check{s}$ek R$\dot{u}$$\check{z}$i$\check{c}$ka, and Konstantin G. Zloshchastiev, Symetry, (2017), {\bf 9}, 165, and references therein.

\end{itemize}

\end{document}